\documentclass{aa}

\usepackage{graphicx}
\usepackage{natbib}

\newcommand{\beq}{\begin{equation}} 
\newcommand{\bea}{\begin{eqnarray}} 
\newcommand{\eeq}{\end{equation}} 
\newcommand{\eea}{\end{eqnarray}} 
\begin{document} 
\title{Cyclotron-Synchrotron: harmonic fitting functions in the 
non-relativistic and trans-relativistic regimes} 

\author{A. Marcowith\inst{1} \and J. Malzac \inst{2,3}}
\offprints{A. Marcowith} 
\institute{{C.E.S.R. Toulouse France \email{Alexandre.Marcowith@cesr.fr}}
\and {Institute of Astronomy, Madingley Road, Cambridge, CB3 0HA, United
Kingdom} \and {Osservatorio di Brera, Via Brera, 28, 20121, Milano, Italy\\
\email{malzac@ast.cam.ac.uk}}} 
 
\date{Received;Accepted} 
\abstract{The present work investigates the calculation of absorption and emission cyclotron  
    line profiles in the non-relativistic and trans-relativistic regimes. We provide fits  
    for the ten first harmonics with synthetic functions down to $10^{-4}$ of the maximum flux with  
    an accuracy of 20 \% at worst. The lines at a given particle energy are calculated from the 
    integration of the Schott formula over the photon and the particle solid angles relative 
    to the magnetic field direction. The method can easily be extended to a larger number  
    of harmonics. We also derive spectral fits of thermal emission line plasmas at non-relativistic 
    and trans-relativistic temperatures extending previous parameterisations. 
   } 

\titlerunning{Cyclotron-Synchrotron: harmonic fitting functions...}
\authorrunning{Marcowith \& Malzac}

\maketitle 
\section{Introduction} 
The cyclotron-synchrotron radiation is produced by charged particles spiraling in a magnetic field. 
The non-relativistic limit (for particle speed $\beta = v/c \ll 1$) is the {\it cyclotron} radiation. 
The relativistic limit (for particle speed $\beta = v/c \sim 1$) is the {\it synchrotron} radiation. 
At intermediate speeds, the process is sometimes called {\it gyro-magnetic} (for $\beta = v/c \sim 0.5$).\\ 
The cyclotron-synchrotron effect is one of the most important processes in astrophysics and has been invoked 
in solar, neutron star physics and in galactic or extra-galactic jets. At non relativistic and 
ultra relativistic limits, the particle emissivity is provided by well known analytical formul{\ae}.\\ 
The main contribution to the cyclotron radiation comes from low harmonics. In most astrophysical 
objects, this radiation is either self-absorbed or absorbed by the environment. It can even 
be suppressed by plasma effects such as in the Razin-Tsytovich effect. However, as stressed 
by different authors (Ghisellini \& Svensson 1991, hereafter GS91; Mason 1992; Gliozzi et al 1996) the cyclotron cross-section
in the non- and mildly relativistic regimes is several orders of 
magnitude above the Thompson cross section. The cyclotron mechanism appears then as a very efficient 
thermalisation process in hot plasmas surrounding compact objects. At higher energies, the synchrotron photons  
are a supplementary source for the Comptonisation process in accretion
discs coron{\ae}, see e.g. Di Matteo et al (1997), Ghisellini, Haardt \&
Svensson (1998), hereafter GHS98, Wardzi\'nski \& Zdziarski (2000), hereafter WZ00 and Wardzi\'nski \& Zdziarski (2001).\\
A detailed numerical investigation of the thermalisation process in the presence of cyclo-synchrotron photons 
has scarcely been discussed (see however GHS98) but is of great importance for the spectral and temporal 
modeling of the complex spectral energy distribution produced in compact objects. Low temperature plasmas 
that may be found in Gamma-ray bursts (in the comoving frame) and in accretion discs may produce cyclotron 
lines combined with synchrotron signatures. Such lines features can be produced in neutron star magnetospheres
probing the magnetic field.\\  
The present work aims to propose useful fitting formul{\ae} for the first ten harmonics dominating  
the emissivity for particle velocities (or normalised momentum) between $\beta = 0.01 \ \rm{and}  
\ p=\beta \gamma = 2$ (see figure 2 in Mahadevan, Narayan \& Yi 1996 hereafter MNY96). We consider the case of 
an isotropic electron distribution in a tangled magnetic field. The method can be extended to an arbitrary number of 
harmonics. At higher energies, however, the  synchrotron formul{\ae} are worth being used. The derived functions 
can then be easily inserted into radiative transfer codes.\\ 
The article is organised as follows: Sect.~\ref{sec:basis} recalls the derivation of cyclo-synchrotron  
emissivity. Sect.~\ref{sec:procedures} deals with the procedures used to compute the emission and the absorption coefficient:  
an integration over a small frequency interval around the cyclotron resonant frequency and 
a direct integration over the angle between the photon and the magnetic  
field. These procedures are tested in Sect.~\ref{sec:tests}. Sect.~\ref{Sntfit} provides  
the fitting functions in the non-thermal case, at a given particle energy. Sect.~\ref{sec:thermem} provides  
the fitting functions for the emission coefficients produced in a thermal plasma.

\section{Cyclo-synchrotron emissivity} \label{sec:basis}
The cyclo-synchrotron spectrum produced by one particle of mass $m$, charge $e$ and velocity  
$v = \beta . c$ embedded in a uniform magnetic field B can be derived  
from classical electrodynamics (see for example Bekefi 1966, chapter 6).
The emitted power spectrum, the so-called {\it Schott formula} is expressed as a sum of harmonics  
peaking at a frequency  $\nu = n \ \nu_{b}/\gamma$  ($n \ge 1$), where $\gamma = (1-\beta^2)^{-1/2}$ is the Lorentz factor
of the particle. The cyclotron frequency is $\nu_b = e \ B / (2 \pi \ m c)$, where the magnetic field B is expressed
in Gauss units. 
\bea 
\frac{dP}{d\nu d\Omega}&&=  2 \pi \frac{e^2}{c} \nu^2 \nonumber \\
&& \sum_{n = 1}^{\infty}
\left[\frac{(\cos\theta-\beta_{\parallel})^2}{\sin^2\theta}
J_n^2(x) + \beta^2_{\perp} J_n'^2(x)\right] \delta(y) \ , 
\label{harmonics} 
\eea 
with $x = \nu \ \gamma \beta_{\perp} \ \sin\theta/\nu_b$, $\beta_{\parallel} = \beta \cos\theta_p$,
$\beta_{\perp} = \beta \sin\theta_p$ and $J_n,J'_n$ being the Bessel functions of (integer) order n and its first derivatives.\\
The cyclotron resonance is defined by  
\beq 
y \ = \ n \frac{\nu_b}{\gamma} - \nu(1 - \beta_{\parallel} \cos\theta) \ = 0 \ . 
\label{reson} 
\eeq 
Eq. (\ref{harmonics}) is expressed in CGS units (e.g. $\rm{erg s^{-1} Hz^{-1} st^{-1}}$) as well as all equations in the text.\\
The angles are defined relative to the dominant magnetic field direction: $\theta_p$ is the particle pitch-angle and $\theta$ is the  
photon angle.\\
Integrating Eq. \ref{harmonics} over the photon angle and the frequency, averaging the pitch-angle leads to
the total power emitted by one particle
\beq
P_{tot} \ = \ \frac{4}{3} \sigma_T \ c \ (\gamma \beta)^2 \ \frac{B^2}{8 \ \pi} \ .
\label{ptot}
\eeq
where the Thomson cross section is $\sigma_T \simeq 6.65 \ 10^{-25} \ \rm{cm^2}$.\\ 
\noindent At the non-relativistic limit, only the first harmonics contribute significantly to the emissivity.  
For the particular pitch-angle $\theta_p = \pi/2$, and after
integration over all observer angles, the power emitted
in a given harmonic $n$ is (Bekefi (1966)) 
\beq 
\eta_n \ = \ \frac{8 \pi^2 \ e^2 \nu_b^2}{c} \ \frac{(n+1)(n^{2n+1})}{(2n+1)!} \ \beta^{2n} \ . 
\label{nonrelat} 
\eeq 
The ratio of two successive harmonics scales roughly as $\beta^2$.\\ 
\noindent In the relativistic limit ($\beta \rightarrow 1$) the emissivity is dominated by the harmonics  
of order $n \sim \gamma^2$. The spectrum tends towards a continuum as the frequency interval between two harmonics  
is $\nu_{b}/\gamma$. The synchrotron power radiated by one particle per unit frequency interval is given by  
(we still assume $\theta_{p} = \pi/2$) 
\bea 
\frac{dP}{d\nu}  \ & = & \  \frac{\sqrt{3} e^{2} \ \nu_{b}}{c} \ 
\frac{\nu}{\nu_{c}} \int_{\frac{\nu}{\nu_{c}}}^{\infty} 
K_{5/3}(t) \ dt \label{powerones} \\  
        \ & \sim & \  \frac{\sqrt{3} e^{2} \ \nu_{b}}{c} \ 1.8 \   
x^{0.3} \ \exp(-x) \ , \nonumber 
\eea 
where the critical frequency is $\nu_{c}= 3/2 \ \nu_{b} \gamma^{2}$, $x \ = \  
\nu/\nu_{c}$ and $K_{5/3}(t)$ is the modified Bessel function of order 5/3. The particle emissivity scales 
as $\nu^{1/3}$ for $x \ll 0.29$ (0.29 is the energy of maximum emissivity) and is exponentially cut-off at high energies 
(Ginzburg \& Syrovatskii (1969)). 

\section{Calculation procedures} \label{sec:procedures} 
We developed two methods to numerically estimate the cyclotron emissivity and absorption cross-section 
of one particle at a given energy and frequency. Each method aims to treat the resonant term inside the Dirac function 
in a correct way, see Eqs. (\ref{harmonics}) and (\ref{reson}). 
 
\subsection{Frequency integration} 
The first procedure consists in integrating the expression (\ref{harmonics}) 
over $\nu$ in a narrow frequency range $[\nu - \Delta \nu/2, \nu + \Delta \nu/2]$, 
with $\Delta \nu = \alpha \ \nu \ll \nu$, $\alpha$ being a positive real variable.\\ 
Eq. (\ref{harmonics}) becomes  
\bea 
\frac{dP}{d\nu \ d\Omega}(\theta_p) & \sim & \frac{2\pi e^2
\nu_b^2}{\alpha \ \nu \ c} \ \sum_{n=n_1}^{n_2}  
\ \frac{n^2}{\gamma^2 (1 - \beta_{\parallel} \cos\theta)^3} \nonumber \\ 
& & \left[\left(\frac{\cos\theta-\beta_{\parallel}}{\sin\theta}\right)^2 \ J_n^2(x_r) +  
\beta^2_{\perp} J_n'^2(x_r)\right]. 
\label{harmonics2} 
\eea 
The resonant term is $x_r = n \beta_{\perp} \ \sin\theta/ (1- \beta_{\parallel} \cos\theta)$.  
The limits $n_1$ and $n_2$ are fixed by the resonant condition (\ref{reson}) taken at $\nu - \Delta \nu/2$ and 
$\nu + \Delta \nu/2$. 
 
\subsection{Direct integration} \label{sec:di}
The second procedure eliminates the resonance through a direct integration over $\theta$, the angle between the photon  
and the magnetic field. The resonance condition selects an angle for a given harmonic $n$, a given pitch-angle and $\beta$ 
\beq 
\cos\theta_r = \frac{1-\frac{n \nu_b}{\gamma \nu}}{\beta_{\parallel}}
\ .
\label{eq:cond} 
\eeq 
Integrated over the solid angle, Eq. (\ref{harmonics}) leads to 
\bea 
\frac{dP}{d\nu}(\theta_p)&&= \frac{2\pi e^2 \nu}{\beta_{\parallel} c}
\nonumber \\
&& \sum_{n=n_1}^{n_2}  
\ \left(\frac{\cos\theta_r-\beta_{\parallel}}{\sin\theta_r}\right)^2 J_n^2(x_r) + \beta^2_{\perp} J_n'^2(x_r) \ . 
\label{harmonics3} 
\eea 
The resonant term is now $x_r= \nu \ \gamma \beta_{\perp} \
sin\theta_r/\nu_b$.
The sum boundaries $n_1$ and $n_2$ define the range of $n$
for which $|\cos\theta_r|$$\leq$1 in Eq.~\ref{eq:cond}.

We note that for $\cos \theta_p$=0,  $\cos\theta_r$ becomes undefinite and
Eq.~(\ref{harmonics3}) breaks down. In this case, all the power is
radiated at the resonant frequencies,  i.e. $dP/d\nu(\theta_p=\pi/2)=0$ except at 
frequencies $\nu=m\nu_b/\gamma$ and at those frequencies the power per unit frequency is actually infinite.
The total power emitted in harmonic $n$ at pitch angle $\theta_p=\pi/2$
may be estimated by integrating Eq.~(\ref{harmonics}) over frequencies around the resonance (leading to a result
similar to Eq.~\ref{harmonics2}) and then numerically over the
observer angles $\theta$.

\subsection{Discussion: total line emission} 
\begin{figure} 
\centering 
\includegraphics[width=\columnwidth]{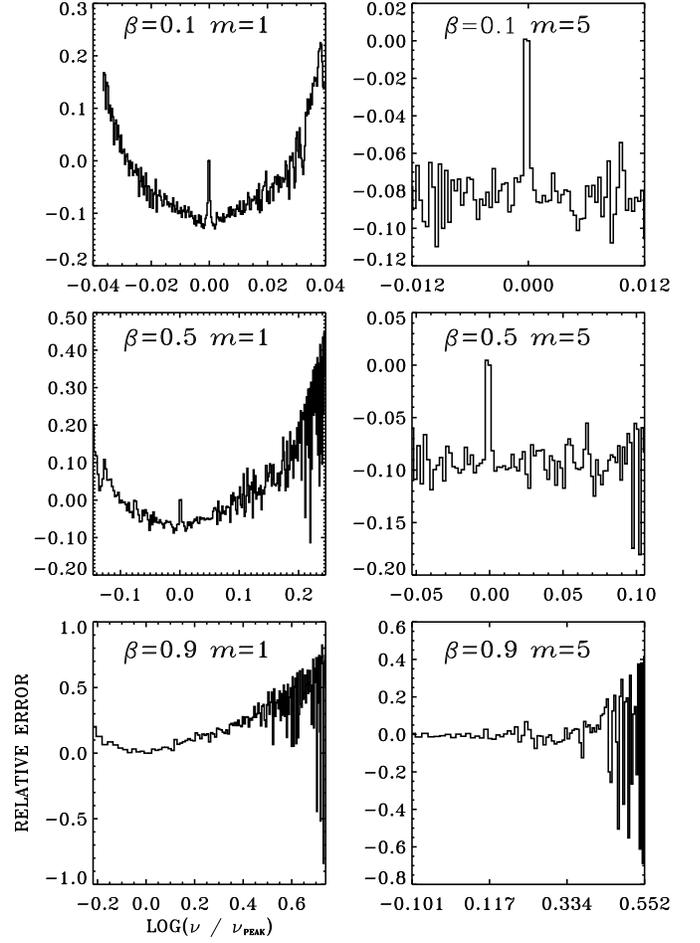} 
\caption{Relative differences between the two formulae (\ref{harmonics2}) and (\ref{harmonics3}) deriving the power emitted 
by one particle. The energy is scaled to the line peak energy. The harmonics 1 and 5 are shown at three particle energies 
$\beta = 0.1, 0.5, 0.9$ corresponding to the energy ranges explored in this work}. 
\label{Fmeth} 
\end{figure}

The power spectrum $dP/d\nu|_n$ at a given harmonic is obtained by
integrating numerically Eq. (\ref{harmonics2}) over the photon solid angle $\Omega$ and over the particle  
pitch-angle $\Omega_p$ using an extended Simpson formula. For the direct integration  
in Eq. (\ref{harmonics3}) only one integral over the particle
pitch-angle is necessary.\\ 
Both methods fail to account for the unavoidable resonance at
frequencies $\nu=n\nu_b/\gamma$ that is due to the contribution from pitch
angle $\theta_p=\pi/2$.
In the frequency integration method, the Dirac function is
artificially broadened over the integration bin of width $\alpha\nu$.
This leads to an overestimate the line emissivity around the resonances,
but the resulting frequency integrated power is conserved.
In the direct integration scheme we neglected the resonant contribution 
at frequencies $\nu=n\nu_b/\gamma$. Although this procedure gives more accurate line profiles,
the frequency integrated power is slightly underestimated.
The two independent methods are compared for different particle energy regimes in Fig.~\ref{Fmeth}.  
They appear to be in agreement within $\sim 10\%$ at low energies and $\sim 20\%$ at high 
particle energies. The errors (the relative difference between the
two emissivities) have been calculated down to $10^{-2}$ the line maximum. The calculated flux
obtained from the second method appears to be lower around the peak of emission at least for the first 
few harmonics and at low particle energies. The discrepancy is always of the order of $10\%$, less 
than the accuracy we aim to reach using the fitting functions. For this reason, we will only 
consider the first (frequency integration) method for the tests detailed in Sect.~\ref{sec:procedures}. 
For both methods, the numerical integration of the power spectrum over frequencies gives the total power (Eq. \ref{ptot}) 
within a few percent.

The best agreements (between the two methods) were obtained using 
an $\alpha$ coefficient (associated with the frequency integration method) in the range $10^{-3}$ to a
few times $10^{-2}$. These values ensure a stable numerical integration. MNY96 have developed a similar but  
different method, approximating the Dirac peak using a smooth function over a frequency interval $\alpha \nu$. 
They reach a good numerical stability with $\alpha = 5 \ 10^{-2}$. In the following, we use the first  
method to find the fitting functions at a given particle energy (see Sect.~\ref{Sntfit}), the second method has the advantage  
of avoiding one integration and will be used in the thermal plasma case (see Sect.~\ref{sec:thermem}). 
 
\subsection{The cyclo-synchrotron absorption cross section} 
The cyclo-synchrotron absorption cross section has been derived by GS91 from the Einstein  
coefficients in a three level system. The differential cross section is the difference  
between true absorption and stimulated emission cross sections.
In the relativistic limit, the stimulated effect takes over for large photon angles with 
the magnetic field, but the total cross section is always positive.\\ 
If the photon energy is much smaller than the electron kinetic energy ($\nu \ll \gamma m_e c^2/h)$, 
the angle integrated cross section can be expressed by the angle
averaged emissivity $j(\nu,\beta)=(dP/d\nu)/4\pi$ in a differential form as 
\beq 
\sigma(\nu,\beta) = \frac{1}{2\ m_e\nu^2} \frac{1}{\gamma \ p} \frac{\partial \ [\gamma \ p  
\ j(\nu,\beta)]}{\partial \gamma} \ . 
\label{crosssect} 
\eeq 
The cyclo-synchrotron cross section can reach values $\sim 10^{16} \sigma_T/B_G$  
for $\nu = \nu_b$, and $\beta \rightarrow 0$.\\
The absorption cross section {\it at a given frequency for a given harmonic} ($n_0 \le 10$, $\beta$ and $\nu$  
are fixed) is calculated using Eq. (\ref{harmonics2}). While integrating over the solid angles, 
we first test if the harmonic $n_0$ is in the range of the permitted harmonics evaluated at 
a given pair of angles $\theta,\theta_p$. The harmonic number interval is defined from Eq.(\ref{reson})
by
\bea
n_1 & = & E[(1-\alpha)\ \nu \ \gamma \ (1-\beta_{\parallel} cos(\theta))] + 1 \ , \\ \nonumber
n_2 & = & E[(1+\alpha)\ \nu \ \gamma \ (1-\beta_{\parallel} cos(\theta))] \ . 
\eea
Once the test is successfully passed, we calculate the analytical derivative over $\gamma$ of the emissivity
at the harmonic number $n_0$, e.g. the derivative of
\bea 
\frac{dP}{d\nu \ d\Omega}|_{n_0} & \sim & \frac{2\pi e^2
\nu_b^2}{\alpha \ \nu \ c} \ \frac{n_0^2}{\gamma^2 (1 - \beta_{\parallel} \cos\theta)^3} \nonumber \\ 
& & \left[\left(\frac{\cos\theta-\beta_{\parallel}}{\sin\theta}\right)^2 \ J_{n_0}^2(x_{r}) +  
\beta^2_{\perp} J_{n_0}'^2(x_{r})\right], 
\label{harmonics0} 
\eea 
with $x_{r} = n_0 \beta_{\perp} \ \sin\theta/ (1- \beta_{\parallel} \cos\theta)$. We finally sum at each
frequency bin the contribution from the ten first harmonics to get $\sigma(\nu,\beta)$.

\section{Tests}\label{sec:tests} 
We performed a series of tests for our numerical schemes in both mono-energetic and thermal cases.  
The methods developed are not well adapted to the relativistic regime as the number of harmonics to  
be summed increases radically when $\beta \rightarrow 1$. They failed to converge to the  
correct analytical expression (\ref{powerones}) at high frequencies.\\ 
We first tried to recover the known cyclotron limit, the total power radiated by one particle,  
at a given harmonic and compared it with expression (\ref{nonrelat}). The second test compares  
the numerical cyclotron absorption coefficient at a given harmonic with the non-relativistic  
coefficient provided by GS91. For the thermal plasma case, we recover the cyclotron absorption  
coefficients given by Chanmugam et al (1989) (hereafter C89) and the Kirchoff law. 
 
\subsection{Tests for mono-energetic distribution} 

\subsubsection{Cyclotron limit}
 
\begin{figure} 
\centering 
\includegraphics[width=\columnwidth]{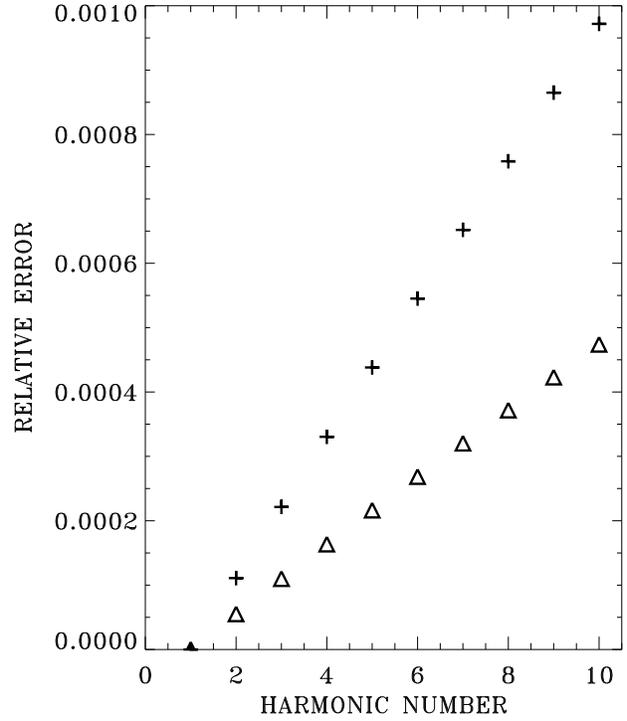} 
\caption{Relative differences between Eq.~(\ref{harmonics2}) the power emitted by one particle integrated over the solid angle 
and Eq.~(\ref{nonrelat}) the power radiated by one particle in the non-relativistic regime. Both emissivities have been calculated
at a pitch-angle $\theta_p = \pi/2$. The plot is provided for 
the ten first harmonics and for two different particle energies $\beta=10^{-2} \ (\rm{triangles}) , \ \rm{and} \ 0.1 \ (\rm{crosses})$ . 
The emissivities have been normalised to the harmonic one. For scaling reasons, the errors have been divided by 50 for  
the $\beta = 0.1$ case.} 
\label{Femnr} 
\end{figure}

In the cyclotron limit ($\beta \rightarrow 0$), we integrate the first harmonics over the resonant  
frequency range with a particle pitch-angle $\theta_p = \pi/2$ and compare our results with Eq.~(\ref{nonrelat}). 
We plot the relative differences for different $\beta$ for the first ten harmonics in Fig.~\ref{Femnr}. As can be seen,  
the agreement is quite good, even if the errors increase with the harmonic number, however they  
are limited to few percent.  

\noindent Concerning the absorption cross section we compared our results with the analytical 
expressions given by GS91. We point out that their formul\ae \ have to be corrected by a factor $1/4 \ \pi$ in 
order to verify Eq.~(\ref{crosssect}). Then, at low energy and at frequencies $\sim \nu_b$, we did obtained
a good agreement and a cross section of the order of $\sigma(\nu,\beta \rightarrow 0) \sim 10^{16} B_G^{-1}  
\sigma_T$. In the trans-relativistic regime, we also found a good agreement at frequencies $\nu \ge \ \nu_b$
(see Fig.~\ref{Fb09}, lower panel).
 
\subsection{Thermal absorption coefficients} 
C89 (and references therein) derived the thermal absorption coefficients for non relativistic thermal  
plasmas (T $\le$ 50 keV) for the two polarisation modes. The ordinary mode has an electric vector field oriented along  
the magnetic field given by the first term in Eq. (\ref{harmonics}), while the extraordinary mode has an electric 
vector field perpendicular to the magnetic field given by the second term in Eq. (\ref{harmonics}). The extraordinary 
mode usually dominates. At different harmonics, we calculate the absorption coefficients from the emissivity at 
a given angle $\theta$ integrated over a Maxwellian distribution $N_e/(\Theta K_2(1/\Theta)) \ 
\gamma^2 \beta \ \rm{exp}(-\gamma/\Theta)$. We have defined the dimensionless temperature as $\Theta = k_B T/ m \ c^2$ and
$N_e$ the electron density.\\ 

\begin{figure}
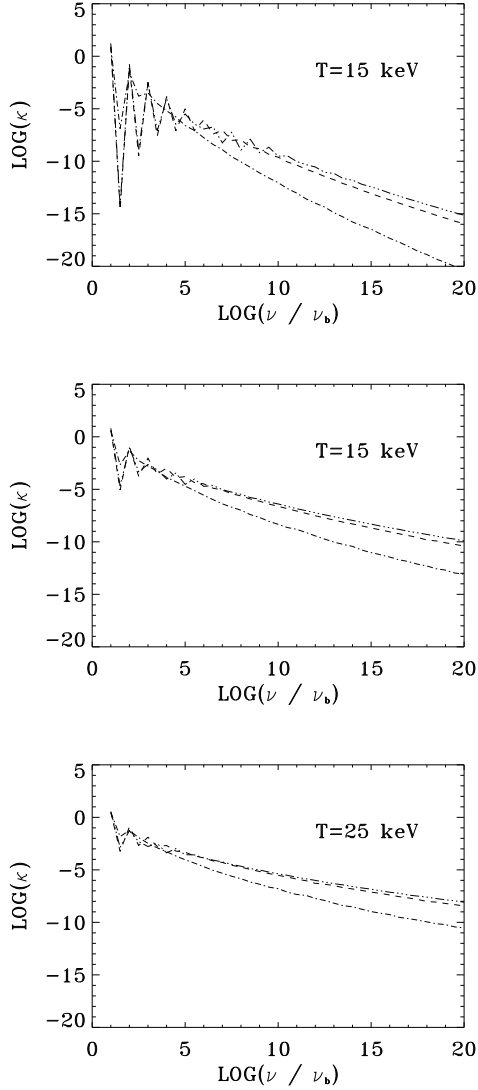
 
\centering 
\scalebox{0.8}{\includegraphics[width=\columnwidth]{3853.f3_1}} 
\scalebox{0.8}{\includegraphics[width=\columnwidth]{3853.f3_2}} 
\scalebox{0.8}{\includegraphics[width=\columnwidth]{3853.f3_3}} 

\caption{Thermal absorption coefficients (in $cm^{-1}$) at $T = 5, 15, 25$ keV. The photon angles 
are $\theta = 80^o \ \rm{three \ dotted-dashed \ line}$ , $60^o \ \rm{long-dashed \ line}$,
$30^o \ \rm{dot-dashed \ line}$. The results can be compared with C89.  
The calculations have been performed for a magnetic field $B = 10 \ \rm{MG}$.} 
\label{Fchan} 
\end{figure}

\noindent We plot our results in figure \ref{Fchan} for three temperatures at three different $\theta$  
values and at different frequencies covering the first 20 harmonics. The agreement with  
C89 is good (see their tables 6A, 6C and 6E), the numerical  thermal coefficient falls close to the extraordinary 
mode values.

\begin{figure} 
\centering 
\includegraphics[width=\columnwidth]{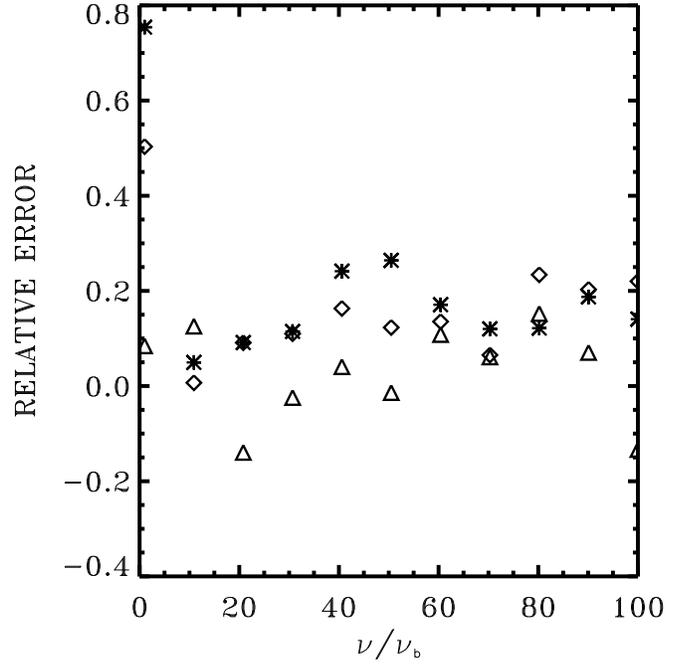} 
\caption{Relative differences between the ratio of the calculated emission and  absorption coefficients $j_{\nu}/\alpha_{\nu}$ 
and  $B_{\nu}(T)$ at non-relativistic temperatures $T = 1 \ \rm{(triangles)}, 10 \ \rm{(squares)}, 50 \ \rm{(stars)}$ keV, in the 
frequency range 1-100 $\nu/\nu_b$.} 
\label{Fkirchoff}
\end{figure}

\noindent Using the same calculation procedure, we check that both thermal emission and absorption coefficients 
fulfill the Kirchoff law, $j_{\nu} = B_{\nu}(T) \ \alpha_{\nu}$. We plot, in figure \ref{Fkirchoff}, 
the relative differences between $j_{\nu}/\alpha_{\nu}$ obtained from numerical integration with $B_{\nu}(T)$ 
versus the frequency for non-relativistic temperatures. The agreement is again good 
except for $T \ge 10$ keV at $\nu \sim \nu_b$ where the flux is weak and the uncertainties obtained with the 
numerical integration method increase. 
 
\section{Non-thermal lines} 
\label{Sntfit}
We now describe the fitting function for the mono-energetic case in the non- and trans-relativistic regimes.  
Whenever it was possible we used {\it simple} (essentially polynomial) functions  
defined by the {\it smallest} number of parameters. Accounting for the evolution of 
the line profile with the particle energy, we split our parameters into two classes: the parameters {\it below}  
the energy of the line peak, the parameters {\it above} the energy of the line peak. The line normalisation with  
$\beta$ and $n$ is also parametrised.\\ 
The generic form for the fitting functions is 
\beq 
F(n,\beta,\nu) = N_1 \ \left(\frac{K_1}{x^{I_1}+K_1-1}\right)^{I_2} \ , 
\label{fitting} 
\eeq 
where  
\bea 
x \ & = & \ \frac{\nu -\nu_{inf}}{\nu_{max}-\nu_{inf}} \ , \ \rm{for \ \nu \le \nu_{max}} \ , 
\\ 
x \ & = & \ \frac{\nu_{sup} -\nu}{\nu_{sup}-\nu_{max}} \ , \ \rm{for \ \nu \ge \nu_{max}} \ . 
\eea 
The different frequencies are $\nu_{inf} = \nu_{b} \ n/(\gamma \ [1+\beta])$, $\nu_{sup} = \nu_{b} 
\ n/(\gamma \ [1-\beta])$ given by the resonant condition Eq.~(\ref{reson}). 
The maximum frequency is $\nu_{max} \ = \nu_{b} \ n/\gamma$. This choice respects 
the normalisation $F(n,\beta,\nu_{max}) = N_1$. The parameters are chosen 
in order for the functions (\ref{fitting}) to fit the line (where the worst error was 20 \%) in a frequency range 
$\nu_{max}$ up (down) to a frequency $\nu_{end \ >}$ ($\nu_{end \ <}$) where the line flux is $10^{-4}$ the peak flux, 
we have the ordering $\nu_{inf} \le (\nu_{end \ <}) \le \nu_{max} \le (\nu_{end \ >}) \le \nu_{sup}$. \\
This value was chosen to ensure good flux estimates at frequencies between two lines maxima  
where in the trans-relativistic regime, two subsequent harmonics may both contribute significantly to the emissivity.\\ 
The parameters are $N_1(n,\beta) \ge 0$ for the line normalisation, $K_1(n,\beta) \ge 0$, 
$I_1(n,\beta) \ge 0$ and $I_2(n,\beta) \ge 0$ for the line profile.\\ 
 
\noindent {\it Once a line is fitted, the $n$ and $\beta$ dependencies of the four  
primary parameters $N_1, K_1, I_1, I_2$ have to be determined using other fitting functions}.  
The whole fitting procedure is twofold, e.g. 
we first derive the previous coefficients in a grid ($\beta,n$) using Eq. (\ref{fitting}). We 
called this the {\bf primary fitting procedure}.  
We then provide {\it new fitting functions} for both $\beta$ and $n$ dependencies of the four previous 
parameters, this is the {\bf secondary fitting procedure}.\\
\noindent Obviously, the general fitting function (\ref{fitting}) is not unique, other  
forms may be much more accurate for the primary fitting procedure but they usually require more parameters and lead 
to a complex second step. The solutions obtained are polynomials of degrees lower than 4, they are accurate 
enough to be used in a wide range of problems dealing with cyclotron radiation.\\

\noindent We consider two energy regimes: the cyclotron-gyro-magnetic (non-relativistic) regime defined  
by $\beta \le 0.5$ and in the trans-relativistic regime defined by $p \le 2$ and $\beta \ge 0.5$.  
Beyond this, the usual relativistic formul{\ae} apply. This choice was motivated by the 
different behavior of the primary parameters in the two regimes (see next).\\ 
For both emission and absorption lines, the second exponent, at a given harmonic,   
$I_2(n)$ has been fixed at $\beta = 0.01$ (the lowest energy considered here) and has an imposed evolution 
in the non relativistic regime (up to $\beta = 0.5$). Above, in the trans relativistic regime, we re-initialise  
$I_2(n)$ to its value at $\beta =0.5$ and impose an other evolution with $\beta$. This procedure limits the  
fluctuations in the other two primary parameters $I_1$ and $K_1$. We also had to consider the evolution  
of the first harmonic individually for both emission and absorption coefficients, and the evolution of the second  
harmonic also individually for the absorption coefficient.\\
\noindent The number of parameters being high (6 parameters for the shape and one for the normalisation), 
the results are unfortunately too complex to be presented in tabular forms. We have therefore decided to provide
fortran routines calculating the power spectrum for frequencies $\nu \ \le 10 \ \nu_b$. 
This procedure has been adopted for the thermal case too (see next section). All the programs are  
accessible via ftp anonymous at {\bf ftp.cesr.fr/pub/synchrotron} (in cyclosynchro.tar.gz). We simply 
present here results in a graphical form. 

\subsection{Results}
\begin{figure}
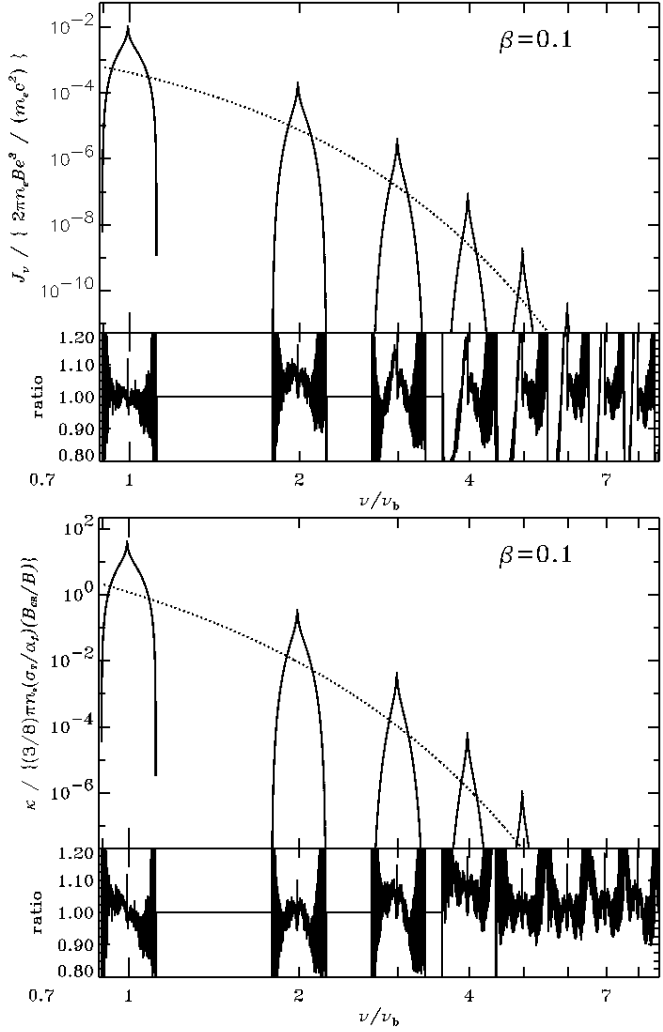
 
\centering 
\scalebox{1}{\includegraphics[width=\columnwidth]{3853.f5_1c}}
\scalebox{1}{\includegraphics[width=\columnwidth]{3853.f5_2c}} 
\caption{Results for non-thermal lines at $\beta = 0.1$. The first upper panel shows the spectra produced using 
the fitting functions (solid line). The flux obtained from numerical integration indistinguishable at this scale has been skipped 
for a better leasibility. The ratio of the fitting function to the numerical results is plotted in the lower panel. 
The second panel shows the fits 
and the ratio for the absorption line. The emission and the absorption coefficients are normalised to $(2 \pi n_e e B^3)/ (m_e c^2)$ and 
$3/8 \ \pi \ N_e \ (\sigma_T/\alpha_f) \ (B_{cr}/B)$ respectively ($\alpha_f = 1/137$ is the fine structure constant 
and $B_{cr}\sim 4.4 \ 10^{13} \ \rm{Gauss}$ is the critical magnetic field). The same normalisation is adopted for all figures. The formula (8) of GHS98 and (2.23) of GS91 in the non-relativistic limit has been used to plot the emission and absorption 
coefficients in dotted line.} 
\label{Fb01} 
\end{figure}

\begin{figure}
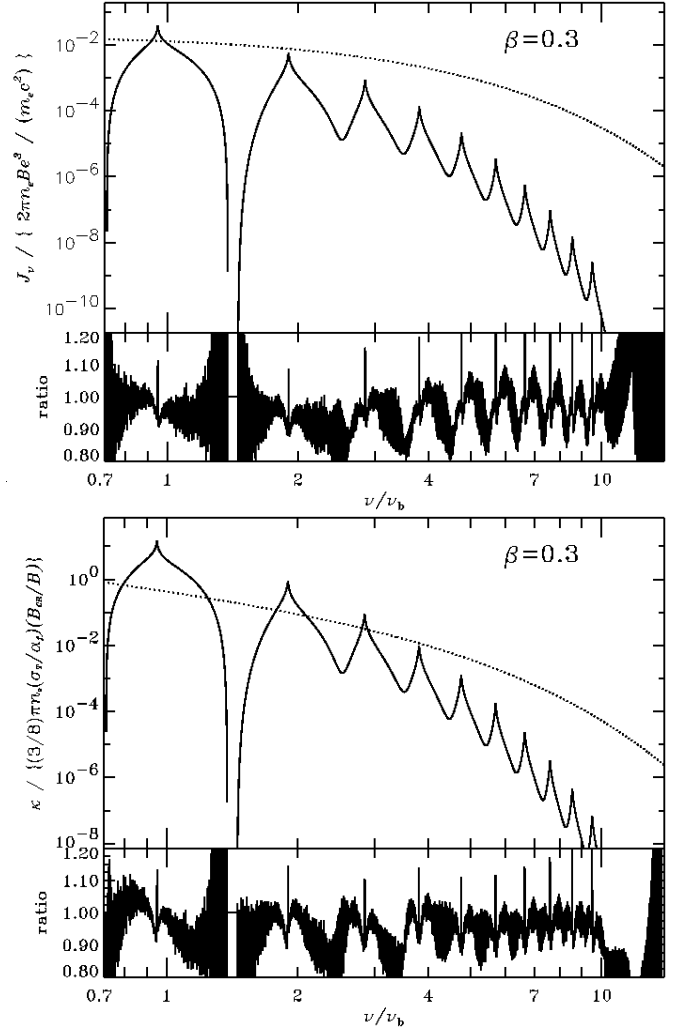
 
\centering 
\scalebox{1}{\includegraphics[width=\columnwidth]{3853.f6_1c}} 
\scalebox{1}{\includegraphics[width=\columnwidth]{3853.f6_2c}} 
\caption{Same as in figure \ref{Fb01} but for $\beta = 0.3$. The formula (2.16) of GS91 in the relativistic limit for the emission 
coefficient is plotted in dotted line in the upper panel. The relativistic formula (2.17) of GS91 in the relativistic limit 
for the absorption cross-section has been used to derive the absorption coefficient plotted in dotted line in the lower panel.} 
\label{Fb03} 
\end{figure}

\noindent For each figure, the first plot presents the numerical calculation and their fits superimposed and the 
second plot presents the relative errors. Figures \ref{Fb01} and \ref{Fb03} display the results for both emission 
and absorption lines in the non-relativistic limits at particle energies corresponding to $\beta = 0.1$ and $\beta =0.3$.  
Figures \ref{Fb07} and \ref{Fb09} display the results from the trans-relativistic regime at particle energies  
corresponding to $\beta= 0.7$ and $\beta = 0.9$.

The units are $\rm{cm^{-1}}$ and $\rm{erg \ s^{-1} \ Hz^{-1} \ cm^{-3} \ st^{-1}} $ for the absorption and emission coefficients 
respectively.\\

In most of the cases, the harmonics are fitted down to $10^{-4}$ the peak line flux with errors never exceeding 20 \% 
(ratio larger than 1.25 or lower than 0.8). Exceptions may be found in figures \ref{Fb01} and \ref{Fb03} for frequencies
$\nu \ge \nu_{max}$ at harmonic number $n \ge 4$ where the profile is very sharp. The flux is overestimated
here but for a restrained frequency range. However, even in these cases errors are around 20 \% and the higher harmonics 
do not strongly contribute to the overall flux. The errors increase at the 
edges of a given harmonic in the non-relativistic regime and at frequencies $\nu$ of the order $\nu_{inf}$ or $\nu_{sup}$ 
in the trans-relativistic regime due to the low accuracy. The noise appearing at the edge of the frequency domain 
(1-10 $\nu_b$) is due to the numerical integration method. As the flux decreases, the method would have 
required a frequency dependent (increasing) $\alpha$ in Eq. (\ref{harmonics2}) to keep a smooth line profile. As 
the flux at these energies are far beyond $10^{-4}$ the peak line flux, we disregarded the problem. However the 
emission (or absorption) beyond $10 \ \nu_b$ is not well reproduced since at these energies harmonics of higher order 
contribute and the synchrotron formula is worth to be used. Finally note that the flux in the absorption of the first 
harmonic in the trans-relativistic case is under-estimated by a factor two under $\nu_{max}$ for fluxes lower 
than $10^{-2}$ the peak flux.\\
Also presented are the non-relativistic and the relativistic limits derived by GHS98 and GS91 respectively. In figure \ref{Fb09}
one can see a good agreement at frequencies $\nu >  \nu_b$ between the numerical and the analytical calculations, for both 
emission and absorption coefficients.\\

\begin{figure} 
\centering 
\scalebox{1}{\includegraphics[width=\columnwidth]{3853.f7_1c}} 
\scalebox{1}{\includegraphics[width=\columnwidth]{3853.f7_2c}} 
\caption{Same as in figure \ref{Fb03} but for $\beta = 0.7$.} 
\label{Fb07} 
\end{figure} 

\begin{figure} 
\centering 
\scalebox{1}{\includegraphics[width=\columnwidth]{3853.f8_1c}} 
\scalebox{1}{\includegraphics[width=\columnwidth]{3853.f8_2c}} 
\caption{Same as in figure \ref{Fb03} but for $\beta = 0.9$.}
\label{Fb09} 
\end{figure} 

\begin{figure}
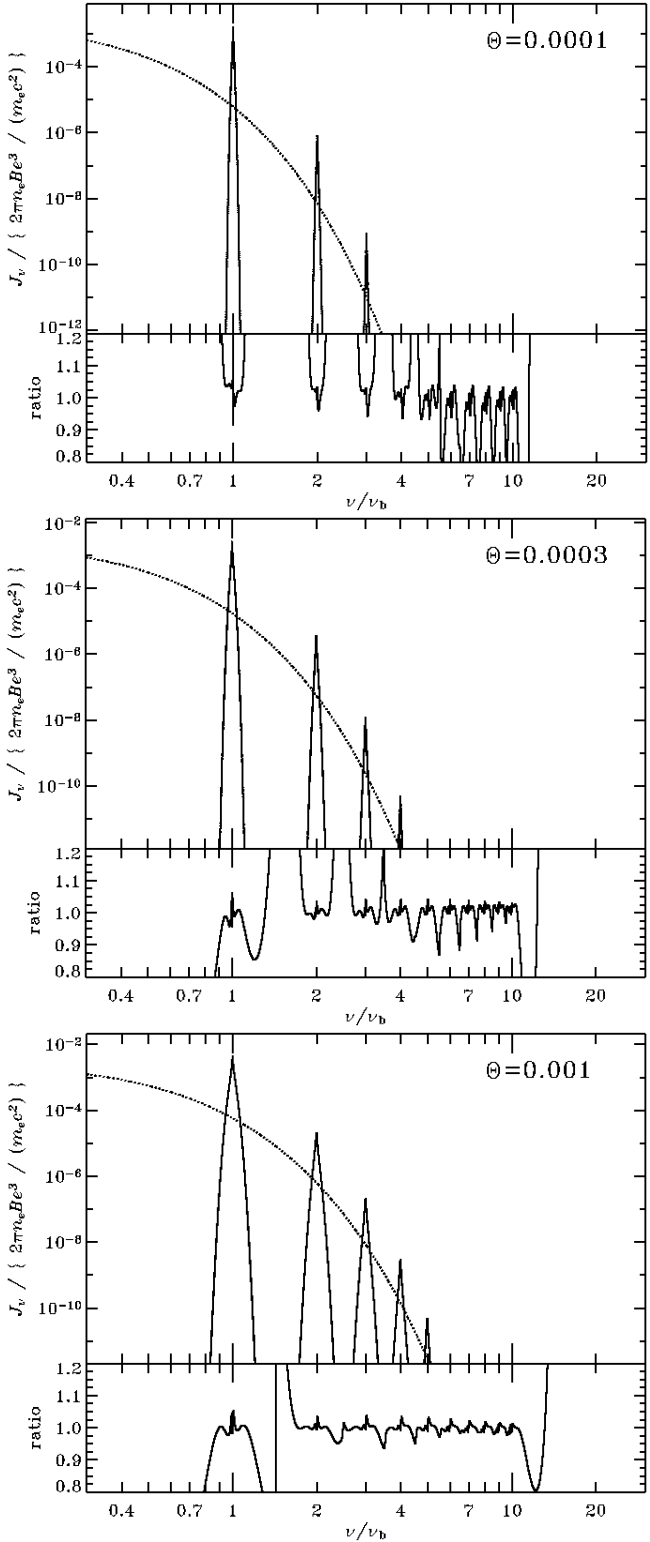
 
\centering 
\scalebox{1}{\includegraphics[width=\columnwidth]{3853.f9_1c}}
\scalebox{1}{\includegraphics[width=\columnwidth]{3853.f9_2c}}
\scalebox{1}{\includegraphics[width=\columnwidth]{3853.f9_3c}}

\caption{From top to bottom: Thermal synchrotron emission coefficients for $kT_{e}$=5.36 10$^{-2}$, 
0.153 and  0.511 keV. Upper panels: the solid curves show the approximation using the fitting functions described 
in Sect.~\ref{sec:thermem}. The results from numerical integration are undiscernable. The dotted curves show the results 
from the formula of WZ00. Lower panels: ratio of the fitting functions to the numerical results.} 
\label{fig:emther1} 
\end{figure}

\begin{figure}
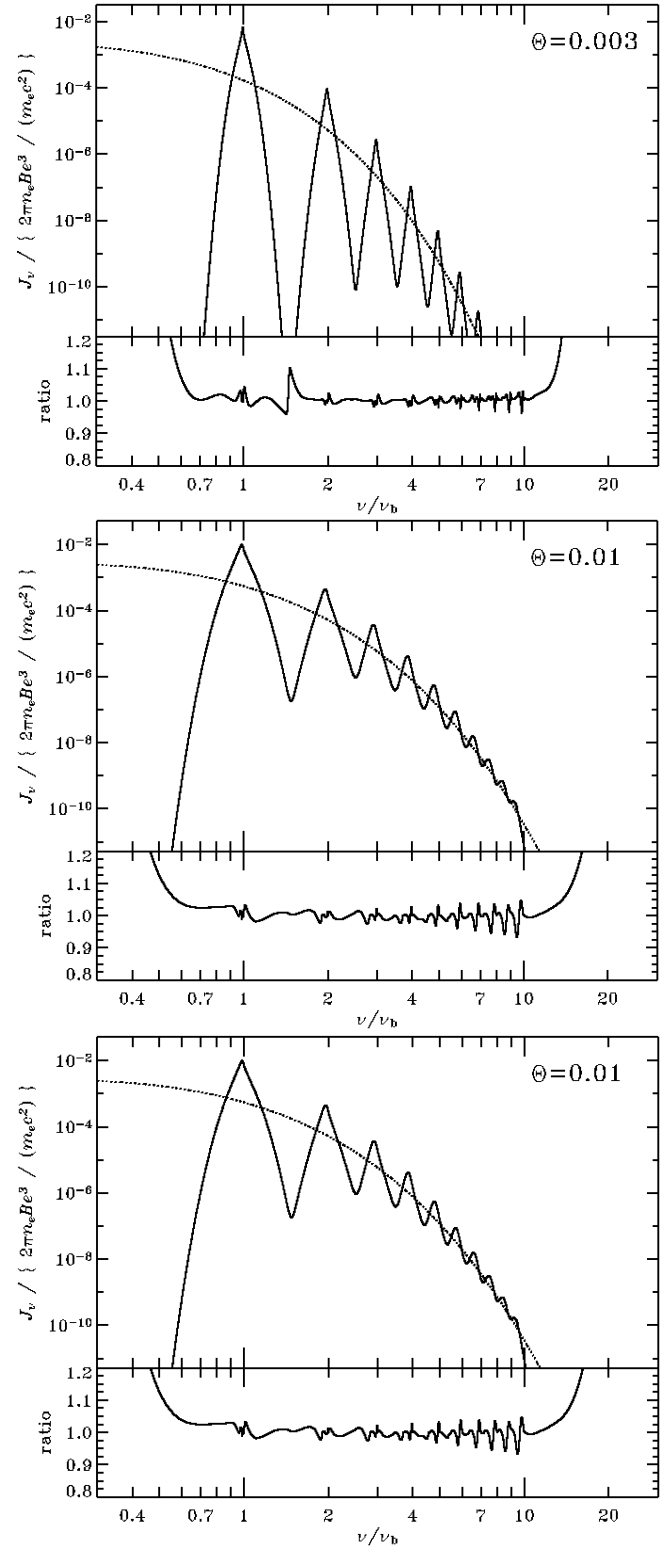
 
\centering 
\scalebox{1}{\includegraphics[width=\columnwidth]{3853.f10_1c}}
\scalebox{1}{\includegraphics[width=\columnwidth]{3853.f10_2c}}
\scalebox{1}{\includegraphics[width=\columnwidth]{3853.f10_3c}}
\caption{Thermal synchrotron emission coefficient for $kT_{e}$=1.53, 
5.11 and  15.3 keV  from top to bottom respectively. See caption of Fig.~\ref{fig:emther1}} 
\label{fig:emther2} 
\end{figure}

\section{Thermal emission lines}
 \label{sec:thermem}
The previous emission coefficients are now integrated over a relativistic  
Maxwellian distribution at a given temperature $\Theta = k_B \ T / m_e c^2$.  
As stated above (Sect.~\ref{sec:di}), for computational facilities, we used the second method 
(direct integration).\\
We computed the line spectra for the ten first harmonics at 200
different temperatures forming a logarithmic grid in the range of
$\Theta = \ 10^{-4}$--$1$. We make available electronically a table of
emissivities in the 10 first harmonics as a function of photon frequency and a
fortran routine reading this table and interpolating between the precomputed emissivities.
The absorption coefficients can then be derived from the Kirchoff law.\\
We found that the emission due to an individual harmonic can be
conveniently approximated as follows: \\
Let $\nu_{max}$ be the photon frequency of the peak of harmonic $n$ (see previous paragraph).
We define the frequencies $\nu_{1}$ and $\nu_{2}$ so that:
\begin{eqnarray}
\nu_{1}&=&\frac{0.999\, \nu_{max}^{2}}{(n+\Theta)\nu_{b}}\\
\nu_{2}&=& (n+0.001)\nu_{b},
\end{eqnarray}
we have $\nu_1 \le \nu_{max} \le \nu_2$.\\
We introduce the following notations:
\begin{eqnarray}
x&=&\nu/(n\nu_{b}),\\
x_{1}&=&\nu_{1}/(n\nu_{b}),\\
x_{2}&=&\nu_{2}/(n\nu_{b}).
\end{eqnarray}
\noindent Let $N_{1}$, $N_{2}$ and $N_{max}$ be the line emissivity
at $\nu_{1}$, $\nu_{2}$ and $\nu_{max}$.
We further note $p_{1}$, $p_{2}$, $p_{3}$... etc, the parameters depending on 
$n$ and $\Theta$ that provide the best fit to our numerical results.
\noindent Then, for $\nu$$<$$\nu_{1}$, the thermal emission coefficient can be
conveniently approximated by the function:
\begin{eqnarray}
F(n,\Theta,\nu)=&N_{1}&\left(\frac{\nu}{\nu_{1}}\right)^{p_{1}}\frac{1-p_{2}x_{1}}{1-p_{2}x} \nonumber \\
&\exp&\left[-\frac{p_{3}}{2\Theta} \left(\frac{1}{x}-\frac{1}{x_{1}}+x-x1\right)\right]
\end{eqnarray}
\noindent For $\nu_{1}$$<$$\nu$$<$$\nu_{max}$:
\begin{equation}
F(n,\Theta,\nu)=N_{max}\exp\left[a\ln(\nu/\nu_{max})^{2}+b\ln(\nu/\nu_{max})^{3}\right]
\end{equation}
\noindent where :
\begin{eqnarray}
a&=&(3-p_{4})\ln(N_{1}/N_{max})/\ln(\nu_{1}/\nu_{max})^{2}\\
b&=&(p_{4}-2)\ln(N_{1}/N_{max})/\ln(\nu_{1}/\nu_{max})^{3}
\end{eqnarray}
\noindent For $\nu_{max}$$<$$\nu$$<$$\nu_{2}$:
\begin{equation}
F(n,\Theta,\nu)=N_{max}\exp\left[c\ln(\nu/\nu_{max})^{2}+d\ln(\nu/\nu_{max})^{3}\right]
\end{equation}
\noindent where :
\begin{eqnarray}
c&=&(3-p_{5})\ln(N_{2}/N_{max})/\ln(\nu_{2}/\nu_{max})^{2}\\
d&=&(p_{5}-2)\ln(N_{2}/N_{max})/\ln(\nu_{2}/\nu_{max})^{3}
\end{eqnarray}
\noindent For $\nu_{2}$$<$$\nu$:
\begin{eqnarray}
F(n,\Theta,\nu)=&N_{2}&\left(\frac{\nu}{\nu_{2}}\right)^{-p_{6}}\frac{x_{2}-1+p_{7}\sqrt{\Theta}}{x-1+p_{7}\sqrt{\Theta}} 
\nonumber \\
&\exp&\left[-\frac{p_{8}}{2\Theta}\left(\frac{1}{x}-\frac{1}{x_{2}}+x-x_{2}\right)\right]
\end{eqnarray}

\noindent This fitting function thus depends on 12 parameters ($\nu_{max}$,
$N_{max}$, $N_{1}$, $N_{2}$ plus 8 parameters $p_{x}$) that were
determined from fits to our tabulated emissivities.
We then constructed a table giving the best fit parameters as a function of 
temperature and harmonic numbers. Contrary to the case of the non-thermal emission,
we have not been able to find simple fitting functions approximating conveniently the temperature dependence 
of these parameters (the second fitting procedure described in Sect.~\ref{Sntfit}).
However, the table of parameters we derived may be used to directly estimate the thermal synchrotron 
emission and absorption coefficients at low photon frequencies. We provide electronically the parameter table 
together with a fortran routine computing the coefficient from that table.
This methods is computationally faster and much less memory consuming
than directly interpolating between 

\clearpage 
\begin{figure}[h]
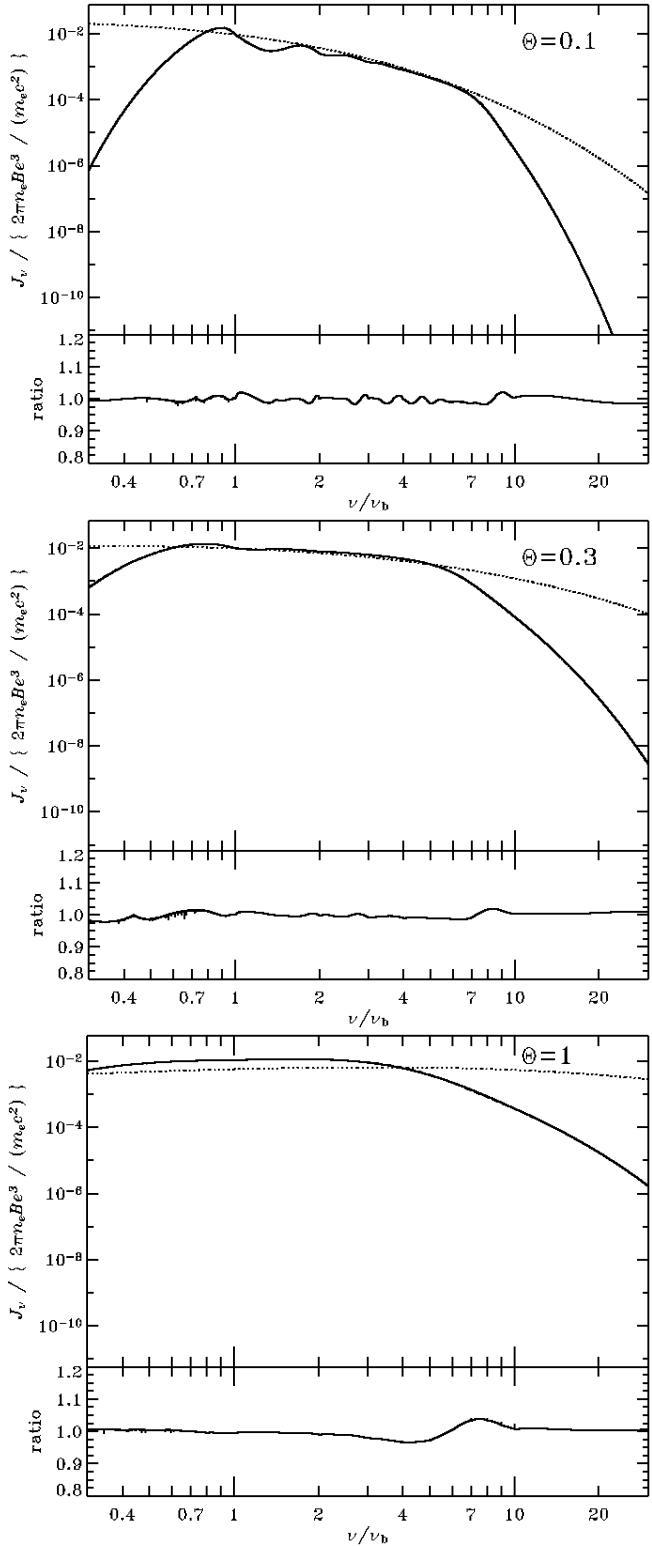
 
\centering 
\scalebox{1}{\includegraphics[width=\columnwidth]{3853.f11_1c}}
\scalebox{1}{\includegraphics[width=\columnwidth]{3853.f11_2c}}
\scalebox{1}{\includegraphics[width=\columnwidth]{3853.f11_3c}}
\caption{Thermal synchrotron emission coefficient for $kT_{e}$=51.1, 
153 and  511 keV from top to bottom respectively. See caption of
Fig.~\ref{fig:emther1}} 
\label{fig:emther3} 
\end{figure} 

the tabulated emissivities. The solutions presented here are much more accurate than the parametrised functions
obtained by different authors (MNY96, WZ00 and references therein) at low temperatures and photon frequencies for which the errors
around $\nu_b$ are particularly severe.

\subsection{Results}\label{sec:results}

Figs.~\ref{fig:emther1}, ~\ref{fig:emther2} and \ref{fig:emther3} 
compare the emission coefficient in $\rm{erg \ s^{-1} \ Hz^{-1} \ cm^{-3} \ st^{-1}}$
obtained using the fitting functions with the results of the numerical integration at different temperatures.
The approximation appears to be good for the whole range of temperatures. 
For illustration we also show in these figures the results of the analytic approximation 
given by WZ00. As long as $\Theta$$<1$ this analytical estimate connects smoothly with our results around
$\nu\sim10\nu_{b}$ and may be used to compute the emission at higher frequencies. On the other hand, 
for $\Theta\sim 1$ and higher, the approximation of WZ00 is not so good, and our approach based on individual 
harmonics becomes inaccurate around $\nu \sim 10\nu_{b}$ due to the significant contribution from higher orders.

\section{Conclusion} 

In this work, we presented two independent methods to compute the cyclo-synchrotron emission and absorption 
coefficients in a tangled magnetic field in the non- and trans relativistic regime. These calculations were 
done both for mono-energetic and thermal electrons. Only the first ten harmonics dominating 
the flux in theses regimes have been considered, but the procedure can easily be extended to a larger 
number of harmonics. We found synthetic fitting functions that reproduce our numerical results with a good accuracy
(with a relative error at worst 20\%). They complement the previously existing approximations 
valid only at higher photon and electron energies. Although these fitting functions are probably too complicated
to be used in analytic calculations, they may prove very useful for numerical purposes.    
For instance, they could be used in high energy radiation transfer codes including both photon and particle 
dynamics (e.g. Coppi 1992; Stern et al. 1995). The fitting functions are made available electronically in the form of
tables and fortran routines that can be easily implemented in such codes. 
 
\acknowledgements 
JM acknowledges financial support from the MURST (grant COFIN98-02-15-41), the European Commission (contract 
number ERBFMRX-CT98-0195, TMR network "Accretion onto black holes, compact stars and protostars"), and PPARC.
The authors thank N.A. Webb for a careful reading of the manuscript.

\bibliographystyle{aa}

\end{document}